\documentclass[11pt,a4paper,american]{extarticle}
\usepackage[T1]{fontenc}
\usepackage[latin1]{inputenc}
\usepackage{amsmath}
\usepackage{amssymb}
\usepackage{setspace}
\usepackage{esint}
\makeatletter

\usepackage{babel}
\usepackage{hyperref}
\usepackage{authblk}
\usepackage{tikz}
\usetikzlibrary{matrix}

\tikzset {_t0b1938qa/.code = {\pgfsetadditionalshadetransform{ \pgftransformshift{\pgfpoint{0 bp } { 0 bp }  }  \pgftransformscale{1 }  }}}
\pgfdeclareradialshading{_7ctm8y52a}{\pgfpoint{0bp}{0bp}}{rgb(0bp)=(1,1,1);
rgb(17.5bp)=(1,1,1);
rgb(24.40476553780692bp)=(0.61,0.61,0.61);
rgb(400bp)=(0.61,0.61,0.61)}
\tikzset{every picture/.style={line width=0.75pt}} 

\usepackage{cancel}
\date{}
\allowdisplaybreaks
\numberwithin{equation}{section}

\author[1]{Thales Azevedo\thanks{thales@if.ufrj.br}}

\affil[1]{Instituto de F\'isica, Universidade Federal do Rio de Janeiro
\authorcr Av. Athos da Silveira Ramos 149, 21941-972, Rio de Janeiro -- Brazil}

\makeatother

\usepackage{babel}
\begin{document}
\title{\bf{Can the Dirac deltas in dipole fields be ignored in classical interactions?}}
\maketitle
\begin{abstract}
When studying (or teaching) classical electromagnetism, one is bound to deal with the electric field of an ideal electric dipole, as well as its magnetic counterpart. A careful analysis then reveals that each of those fields must include, for consistency, a term proportional to a Dirac delta function localized at the position of the dipole. However, one is usually told not to worry about those terms since, as classical interactions always involve sources which are spatially separated, the Dirac-delta terms are only relevant for quantum mechanics, where they are directly related to important phenomena. In this work, we pose and solve a purely classical problem in electrostatics  in which the Dirac-delta terms in the dipole fields are indispensable. It involves the computation of the interaction energy between a conductor with a spherical cavity and an (ideal) electric dipole located at the center of that cavity. We also solve its magnetic counterpart, replacing the conductor with a superconductor and the electric dipole with a magnetic one.

\end{abstract}

\thispagestyle{empty}

\newpage 

\section{Introduction\label{sec:Introduction}}

The importance of studying electric and magnetic dipoles in classical electromagnetism cannot be overstated. Indeed, (ideal) electric dipoles constitute an excellent model for polar molecules, and any polarizable material can be described (macroscopically)  as a continuum of dipoles \cite{Griffiths,Jackson,Zangwill}. Moreover, the interaction between electric dipoles is at the heart of dispersive forces such as van der Waals's \cite{Taddei}. As for the magnetic dipole, it can in many cases be a good approximation
to a real magnet (or electromagnet), and is somewhat fundamental since there is no magnetic monopole in Maxwell's theory. One can also describe any magnetic material as a continuum of ideal magnetic dipoles. 

In order to compute the electric field $\vec{E}_\mathrm{dip}$ of a static electric dipole $\vec{p}$ located at position $\vec{r}_p$, one usually starts from its electric potential
\begin{equation}
V_\mathrm{dip}(\vec{r}\,) = \frac{1}{4\pi\varepsilon_0} \frac{\vec{p}\cdot (\vec{r} - \vec{r}_p)}{|\vec{r} - \vec{r}_p|^3}\,,
\end{equation}
which can be obtained in several different ways, and then one proceeds to calculating the corresponding electric field through $\vec{E}_\mathrm{dip} = -\vec{\nabla}V_\mathrm{dip}$. Now, as a function, it is clear that $V_\mathrm{dip}(\vec{r}\,)$ is not defined at $\vec{r} = \vec{r}_p$. When this fact is carefully taken into account, one finds
\begin{equation}
\vec{E}_\mathrm{dip}(\vec{r}\,) = \frac{1}{4\pi\varepsilon_0} \left[\frac{3\,\Big[\vec{p}\cdot (\vec{r} - \vec{r}_p)\Big](\vec{r} - \vec{r}_p) - (\vec{r} - \vec{r}_p)^2 \vec{p}}{|\vec{r} - \vec{r}_p|^5}\,\right] - \frac{1}{3\varepsilon_0}\delta^3(\vec{r} - \vec{r}_p)\,\vec{p}\,.
\label{Edip}
\end{equation}

The last term in the above expression is the part that does not arise naturally from just taking the gradient of $V_\mathrm{dip}(\vec{r}\,)$. Being proportional to a Dirac delta function, it is nonzero only at the position of the dipole. Hence, in most classical applications, it does not need to be considered.


Similarly, one can compute the magnetic field $\vec{B}_\mathrm{dip}$ of a static magnetic dipole $\vec{m}$ located at position $\vec{r}_m$. The resulting expression is
\begin{equation}
\vec{B}_\mathrm{dip}(\vec{r}\,) = \frac{\mu_0}{4\pi} \left[\frac{3\,\Big[\vec{m}\cdot (\vec{r} - \vec{r}_m)\Big](\vec{r} - \vec{r}_m) - (\vec{r} - \vec{r}_m)^2 \vec{m}}{|\vec{r} - \vec{r}_m|^5}\,\right] + \frac{2\mu_0}{3}\delta^3(\vec{r} - \vec{r}_m)\,\vec{m}\,.
\label{Bdip}
\end{equation}

In this paper, we will not be concerned with the derivation of equations (\ref{Edip}) and (\ref{Bdip}) --- rather, our goal here is to discuss to what extent the delta-function terms in the above expressions can be neglected in classical interactions.  Nonetheless, we remind the reader of the simple argument given in a famous paper by Griffiths for the existence of those delta-function terms  \cite{Griffiths2}. If we think of an ideal electric dipole $\vec{p}$ as  the limit of a uniformly polarized sphere of radius $R$, then the electric field outside the sphere is exactly equal to that of an ideal dipole, whereas the field inside the sphere is uniform,  given by $-(1/4\pi\varepsilon_0)\vec{p}/R^3$. Clearly, the field inside the sphere blows up in the $R\to0$ limit, but the integral of that field over the volume of the sphere gives $-(1/4\pi\varepsilon_0)\vec{p}/R^3\times 4\pi R^3/3 = -(3\varepsilon_0)^{-1}\vec{p}$, no matter how small $R$ is. This suggests that the electric field ``inside'' an ideal dipole can be written precisely as the delta-function term in equation (\ref{Edip}).

 The delta-function term in equation (\ref{Bdip}) can be explained in a similar way, if we think of an ideal magnetic dipole as the limit of a uniformly magnetized sphere. In particular, the relative sign of the magnetic delta-function term with respect to the electric one can be thought of as stemming from the fact that the magnetic field at the center of a small loop of current  points in the same direction as its magnetic dipole moment, whereas the electric field at the midpoint of a (real) dipole has the opposite direction of its electric dipole moment.
 The reader can find more details in any of the books cited in the beginning of this introduction --- see also a very nice review and pedagogical explanation given in ref. \cite{Farina}. 

The paper is organized as follows. In section 2, we briefly review the calculation of the interaction energy of two systems in classical electromagnetism in terms of their electric or magnetic fields, and we give a general argument as to why the delta-function terms should matter for that type of calculation. In section 3, we set up a specific electrostatics problem in which the need for the electric-dipole delta-function term is easy to check. In section 4, we pose and solve an analogous problem for the magnetic case. We conclude with some remarks in section 5.

\section{Brief review of the energy stored in the fields\label{sec:review}}

In section IV of ref.  \cite{Griffiths2},  it is stated that ``In most applications (...) the delta-function term can be ignored.'' Then, the author explains its relevance for quantum mechanics, which might suggest that those terms can be ignored in classical electromagnetism. More recently, in ref. \cite{Farina}, the authors affirm that ``As the Dirac delta function vanishes outside the origin [i.e. the position of the dipole] and the classical interaction is always between sources spatially separated, in pure classical electromagnetism these extra terms play no important role.'' At first, those considerations seem very sensible. However, let us see what we can learn   by revisiting the electromagnetic energy stored in the fields.

It can be shown that the energy density $\cal{E}$ contained in the electric and magnetic fields is given by
\begin{equation}
{\cal{E}} = \frac{1}{2}\left(\varepsilon_0 |\vec{E}|^2 + \frac{1}{\mu_0} |\vec{B}|^2\right).
\end{equation}
In fact, in the introduction of his book \cite{Wald}, Wald argues that from a relativistic point of view the above equation ``should be viewed as having fundamental status in the theory of
electromagnetism, comparable to that of Maxwell's equations.'' Thus, choosing zero energy to mean  the situation in which the   electric and magnetic fields vanish everywhere, the energy $U$ of any electromagnetic system can be written as
\begin{equation}
U =  \frac{1}{2}\int_{\cal{R}}\left(\varepsilon_0 |\vec{E}|^2 + \frac{1}{\mu_0} |\vec{B}|^2\right)\mathrm{d}^3 \vec{r}\,,
\end{equation}
where $\cal{R}$ stands for the whole space.

Now, consider two subsystems, 1 and 2, interacting electrostatically. The electric field $\vec{E}$ can be decomposed into contributions coming from each subsystem, i.e. $\vec{E} = \vec{E}_1 + \vec{E}_2$. In this case, the energy of the system can be written as
\begin{equation}
U = U_1 + U_2 + U_{12}\,,
\end{equation}
where $U_j \equiv \frac{1}{2}\int_{\cal{R}} \varepsilon_0 |\vec{E}_j|^2 \,\mathrm{d}^3 \vec{r}$ corresponds to the self-energy of subsystem $j$ $(j\in\{1,2\})$
and
\begin{equation}
U_{12} \equiv \varepsilon_0 \int_{\cal{R}} \vec{E}_1 \cdot\vec{E}_2\,\mathrm{d}^3 \vec{r}
\label{Uint}
\end{equation}
is the interaction energy of the two subsystems. Note that, if one of the fields contains a delta-function term, then that term will in general contribute to the dot product in $U_{12}$. Evidently, an analogous expression would hold in the case of two subsystems interacting magnetostatically.

In the following section, we will discuss a specific electrostatics problem in which the delta-function contribution is easy to calculate.

\section{Electric dipole in conductor cavity}

We have argued that the delta-function term in the electric field of a point dipole will in general yield a finite contribution to the electrostatic energy of a system containing the dipole.  It would be nice to see explicitly in an example how that contribution appears and has precisely the value which is necessary for consistency. For most configurations the integral in equation (\ref{Uint}) is not easy to compute. However, there is a typical electrostatics problem in which the computation is straightforward.


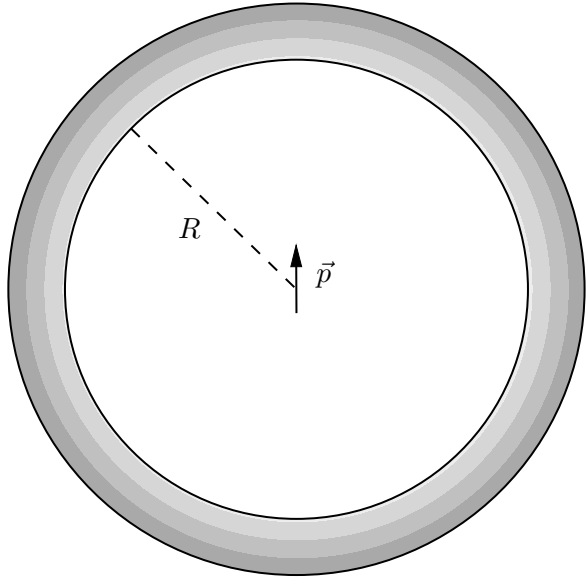
\begin{figure}[h]

\centering

\begin{tikzpicture}[x=0.75pt,y=0.75pt,yscale=-1,xscale=1]

\path  [shading=_7ctm8y52a,_t0b1938qa] (164.75,151.83) .. controls (164.75,72.44) and (229.11,8.08) .. (308.5,8.08) .. controls (387.89,8.08) and (452.25,72.44) .. (452.25,151.83) .. controls (452.25,231.22) and (387.89,295.58) .. (308.5,295.58) .. controls (229.11,295.58) and (164.75,231.22) .. (164.75,151.83) -- cycle ; 
 \draw   (164.75,151.83) .. controls (164.75,72.44) and (229.11,8.08) .. (308.5,8.08) .. controls (387.89,8.08) and (452.25,72.44) .. (452.25,151.83) .. controls (452.25,231.22) and (387.89,295.58) .. (308.5,295.58) .. controls (229.11,295.58) and (164.75,231.22) .. (164.75,151.83) -- cycle ; 

\draw  [fill={rgb, 255:red, 255; green, 255; blue, 255 }  ,fill opacity=1 ] (193,151.83) .. controls (193,88.04) and (244.71,36.33) .. (308.5,36.33) .. controls (372.29,36.33) and (424,88.04) .. (424,151.83) .. controls (424,215.62) and (372.29,267.33) .. (308.5,267.33) .. controls (244.71,267.33) and (193,215.62) .. (193,151.83) -- cycle ;
\draw    (308.5,163.83) -- (308.34,131.67) ;
\draw [shift={(308.33,129.67)}, rotate = 89.72] [fill={rgb, 255:red, 0; green, 0; blue, 0 }  ][line width=0.08]  [draw opacity=0] (10.8,-2.7) -- (0,0) -- (10.8,2.7) -- cycle    ;
\draw  [dash pattern={on 4.5pt off 4.5pt}]  (226,71) -- (308.5,151.83) ;

\draw (248,114.4) node [anchor=north west][inner sep=0.75pt]    {$R$};
\draw (317,136.4) node [anchor=north west][inner sep=0.75pt]    {$\vec{p}$};

\end{tikzpicture}

\caption{An ideal dipole $\vec{p}$ at the center of a spherical cavity of radius $R$ within a (hollow) spherical conductor. In this situation, there is an induced surface charge with density $\sigma$ on the inner surface of the conductor, i.e. the surface of the cavity.}
\label{conductor}

\end{figure}


Consider a globally neutral spherical conductor containing a spherical cavity of radius $R$, as illustrated in figure \ref{conductor}. In that cavity, there is nothing but an ideal electric dipole $\vec{p} = p_0\,\hat{\mathtt{z}}$, located at the center of the cavity, which we take as the origin of our coordinate system. Since the system is in electrostatic equilibrium, there is  an induced charge distribution on the surface of the cavity which ensures that the (total) electric field in the bulk of the conductor is zero. (In fact, since the conductor is neutral, there is no charge on its outer surface, which implies the electric field is vanishing for $r>R$.) Hence, there are two contributions to the electric field, as in the generic case discussed in the previous section. Namely the dipole field 
\begin{equation}
\vec{E}_1(\vec{r}\,) \equiv \vec{E}_\mathrm{dip}(\vec{r}\,) = \frac{1}{4\pi\varepsilon_0} \left[\frac{3(\vec{p}\cdot \hat{\mathtt{r}})\hat{\mathtt{r}} -  \vec{p}}{{r}^3}\,\right] - \frac{1}{3\varepsilon_0}\delta^3(\vec{r}\,)\,\vec{p}\,,
\label{EdipO}
\end{equation}
which is just equation (\ref{Edip}) with $\vec{r}_p = \vec{0\!}$ (since the dipole is at the origin), and the field produced by the  charge distribution on the surface of the cavity
\begin{equation}
\vec{E}_2(\vec{r}\,) \equiv \vec{E}_\sigma(\vec{r}\,) = \frac{1}{4\pi\varepsilon_0} \oint_{S^2} \frac{\sigma(\vec{r}\,{}^\prime)\,\mathrm{d}^2\vec{r}\,{}^\prime}{|\vec{r}-\vec{r}\,{}^\prime|^3} (\vec{r}-\vec{r}\,{}^\prime)\,,
\end{equation}
where $\sigma(\vec{r}\,{}^\prime)$ denotes the surface charge density.

Now, it is well known \cite{Griffiths} that a sphere of radius $R$ with surface charge density of the form
\begin{equation}
\sigma(\vec{r}\,{}^\prime) = C\, \hat{\mathtt{z}}\cdot\hat{\mathtt{r}}^\prime\,,
\end{equation}
where $C$ is a real constant, gives rise to the following electric field:
\begin{equation}
\vec{E}_\sigma(\vec{r}\,) = \left\{ \begin{array}{l}\displaystyle -\frac{C}{3\varepsilon_0} \hat{\mathtt{z}}\,, \qquad r < R  \\   \\  \displaystyle \frac{1}{4\pi\varepsilon_0} \left[\frac{3(\vec{c}\cdot \hat{\mathtt{r}})\hat{\mathtt{r}} -  \vec{c}}{{r}^3}\,\right], \qquad r > R\,,  \end{array} \right.
\end{equation}
where $\vec{c} \equiv \frac{4}{3}\pi R^3C\,\hat{\mathtt{z}}$. Since we are interested in a charge distribution such that the electric field vanishes in the bulk of the conductor, i.e. for $r > R$, we need to take $C = -p_0\left(\frac{4}{3}\pi R^3\right)^{-1}$, which implies $\vec{c} = -\vec{p}$. Therefore, we get 
\begin{equation}
\vec{E}_2(\vec{r}\,) = \left\{ \begin{array}{l}\displaystyle \frac{1}{4\pi\varepsilon_0} \frac{\vec{p}}{R^3}\,, \qquad r < R  \\   \\  \displaystyle -\frac{1}{4\pi\varepsilon_0} \left[\frac{3(\vec{p}\cdot \hat{\mathtt{r}})\hat{\mathtt{r}} -  \vec{p}}{{r}^3}\,\right], \qquad r > R\,.  \end{array} \right.
\end{equation}
We remind the reader that the solution to this electrostatics problem is unique.

We are now ready to calculate the interaction energy of this dipole-conductor system. First, regarding $\vec{E}_2$ as an external field with respect to the dipole, we can apply the usual potential energy formula $U = -\vec{p}\cdot\vec{E}^\mathrm{(ext)}$ to obtain\footnote{Note that the $U = -\vec{p}\cdot\vec{E}^\mathrm{(ext)}$ formula does not directly involve the dipole field, hence the delta-function term plays no role here. Although it can also be  derived from  equation (\ref{Uint}), it has a straightforward interpretation in terms of mechanical work.}
\begin{equation}
U_{12} = \left.-\vec{p}\cdot\vec{E}_2\right|_{r=0} = -\frac{1}{4\pi\varepsilon_0} \frac{p_0^2}{R^3}\,.
\label{energia}
\end{equation}
Now we are going to show that the same result can be obtained from the more fundamental expression in equation (\ref{Uint}), provided that we include the contribution arising from the delta-function term in $\vec{E}_1$. We can split $U_{12}$ in two terms, as follows:
\begin{equation}
U_{12} = \varepsilon_0 \int_{r>R} \vec{E}_1 \cdot\vec{E}_2\,\mathrm{d}^3 \vec{r}\, +\, \varepsilon_0 \int_{r<R} \vec{E}_1 \cdot\vec{E}_2\,\mathrm{d}^3 \vec{r}\,.
\label{UintComp}
\end{equation}
For the integrand in the first term, we have 
\begin{equation}
{{\cal I}}_>^E \equiv \left.\vec{E}_1 \cdot\vec{E}_2\right|_{r>R} = -\left(\frac{1}{4\pi\varepsilon_0}\right)^2 \left[\frac{3(\vec{p}\cdot \hat{\mathtt{r}})\hat{\mathtt{r}} -  \vec{p}}{{r}^3}\,\right]^2 = -\left(\frac{1}{4\pi\varepsilon_0}\right)^2 \frac{3\left(\hat{\mathtt{z}}\cdot\hat{\mathtt{r}}\right)^2 + 1}{r^6}p_0^2\,.
\end{equation}
This function can be straightforwardly integrated  using spherical coordinates ($\hat{\mathtt{z}}\cdot\hat{\mathtt{r}} = \cos\theta$). Then we find
\begin{equation}
\varepsilon_0 \int_{r>R} \vec{E}_1 \cdot\vec{E}_2\,\mathrm{d}^3 \vec{r} = -\frac{1}{6\pi\varepsilon_0} \frac{p_0^2}{R^3}\,.
\label{uE1}
\end{equation}

For the integrand in the second term on the right-hand side of equation (\ref{UintComp}), we have

\begin{equation}
{{\cal I}}_<^E \equiv \left.\vec{E}_1 \cdot\vec{E}_2\right|_{r<R} = \left(\frac{1}{4\pi\varepsilon_0}\right)^2 \frac{\vec{p}}{R^3} \cdot \left[\frac{3(\vec{p}\cdot \hat{\mathtt{r}})\hat{\mathtt{r}} -  \vec{p}}{{r}^3}\,\right]    - \frac{1}{3\varepsilon_0}\delta^3(\vec{r}\,)\left(\frac{1}{4\pi\varepsilon_0}\right) \frac{\vec{p}}{R^3} \cdot\vec{p}\,.
\end{equation}
Note that in this case, as opposed to the computation of ${{\cal I}}_>^E$, we must include the contribution from the delta-function term in $\vec{E}_1$, since here the region of integration includes the origin (where the dipole is located). After integration, the contribution from the first term in ${{\cal I}}_<^E$ vanishes, because it is proportional to $\int_0^\pi [3\cos^2\theta - 1]\sin\theta\,\mathrm{d}\theta = 0$. Then we are left with the contribution from the delta-function term, which is trivial to compute and yields
\begin{equation}
\varepsilon_0 \int_{r<R} \vec{E}_1 \cdot\vec{E}_2\,\mathrm{d}^3 \vec{r} = -\frac{1}{12\pi\varepsilon_0} \frac{p_0^2}{R^3}\,.
\label{uE2}
\end{equation}

Therefore, as expected, the sum of (\ref{uE1}) and (\ref{uE2}) exactly reproduces  equation (\ref{energia}). It should be clear that the contribution stemming from the delta-function term in the electric field of the dipole was crucial for the matching of the two results.

\section{Magnetic dipole in superconductor cavity}

In this section, we solve a problem in magnetostatics which is completely analogous to the electrostatics problem discussed in the previous section. Here the delta-function term in the magnetic field of an ideal magnetic dipole will also play a crucial role in the computation of the interaction energy.

Consider again a system like the one in figure \ref{conductor}, but now with the spherical conductor replaced by a type-I superconductor of the same shape, including the spherical cavity of radius $R$. Moreover, the electric dipole at the center of the cavity is now replaced by an ideal magnetic dipole $\vec{m} = m_0\,\hat{\mathtt{z}}$. Because of the  Meissner effect, the (total) magnetic field in the bulk of the superconductor is zero. This is achieved thanks to an induced current distribution on the surface of the cavity. Hence, there are two contributions to the magnetic field. Namely the dipole field 
\begin{equation}
\vec{B}_1(\vec{r}\,) \equiv \vec{B}_\mathrm{dip}(\vec{r}\,) = \frac{\mu_0}{4\pi} \left[\frac{3(\vec{m}\cdot \hat{\mathtt{r}})\hat{\mathtt{r}} -  \vec{m}}{{r}^3}\,\right] + \frac{2\mu_0}{3}\delta^3(\vec{r}\,)\,\vec{m}\,,
\label{BdipO}
\end{equation}
which is just equation (\ref{Bdip}) with $\vec{r}_m = \vec{0\!}$ (since the dipole is at the origin), and the field produced by the  current distribution on the surface of the cavity
\begin{equation}
\vec{B}_2(\vec{r}\,) \equiv \vec{B}_K(\vec{r}\,) = \frac{\mu_0}{4\pi} \oint_{S^2} \frac{\vec{K}(\vec{r}\,{}^\prime)\,\mathrm{d}^2\vec{r}\,{}^\prime}{|\vec{r}-\vec{r}\,{}^\prime|^3}\times (\vec{r}-\vec{r}\,{}^\prime)\,,
\end{equation}
where $\vec{K}(\vec{r}\,{}^\prime)$ denotes the surface current density.

Now, it is well known \cite{Griffiths} that a sphere of radius $R$ with surface current density of the form
\begin{equation}
\vec{K}(\vec{r}\,{}^\prime) = D\, \hat{\mathtt{z}}\times\hat{\mathtt{r}}^\prime\,,
\end{equation}
where $D$ is a real constant, gives rise to the following magnetic field:
\begin{equation}
\vec{B}_K(\vec{r}\,) = \left\{ \begin{array}{l}\displaystyle \frac{2}{3} \mu_0 D\, \hat{\mathtt{z}}\,, \qquad r < R  \\   \\  \displaystyle \frac{\mu_0}{4\pi} \left[\frac{3(\vec{d}\cdot \hat{\mathtt{r}})\hat{\mathtt{r}} -  \vec{d}}{{r}^3}\,\right], \qquad r > R\,,  \end{array} \right.
\end{equation}
where $\vec{d} \equiv \frac{4}{3}\pi R^3D\,\hat{\mathtt{z}}$. Since we are interested in a current distribution such that the magnetic field vanishes in the bulk of the superconductor, i.e. for $r > R$, we need to take $D = -m_0\left(\frac{4}{3}\pi R^3\right)^{-1}$, which implies $\vec{d} = -\vec{m}$. 
Therefore, we get 
\begin{equation}
\vec{B}_2(\vec{r}\,) = \left\{ \begin{array}{l}\displaystyle -\frac{\mu_0}{2\pi} \frac{\vec{m}}{R^3}\,, \qquad r < R  \\   \\  \displaystyle -\frac{\mu_0}{4\pi} \left[\frac{3(\vec{m}\cdot \hat{\mathtt{r}})\hat{\mathtt{r}} -  \vec{m}}{{r}^3}\,\right], \qquad r > R\,.  \end{array} \right.
\end{equation}

From this point, the math is almost identical to that of the analogous electrostatics problem. The interaction energy of the dipole-superconductor system is given by
\begin{eqnarray}
U_{12} & = & \frac{1}{\mu_0} \int_{r>R} \vec{B}_1 \cdot\vec{B}_2\,\mathrm{d}^3 \vec{r}\, +\, \frac{1}{\mu_0} \int_{r<R} \vec{B}_1 \cdot\vec{B}_2\,\mathrm{d}^3 \vec{r} \nonumber \\
 & = &  -\frac{\mu_0}{6\pi} \frac{m_0^2}{R^3} -\frac{\mu_0}{3\pi} \frac{m_0^2}{R^3} \nonumber \\
 & = & -\frac{\mu_0}{2\pi} \frac{m_0^2}{R^3}\,.
\label{UintCompB}
\end{eqnarray}
It is straightforward to check that the same result can be obtained from the formula $U = \vec{m}\cdot\vec{B}^\mathrm{(ext)}$.\footnote{Note the absence of a minus sign here, as opposed to the electrostatic analogue $U = -\vec{p}\cdot\vec{E}^\mathrm{(ext)}$. A thorough discussion of that point is somewhat subtle and out of the scope of this paper. However, note that $U = \vec{m}\cdot\vec{B}^\mathrm{(ext)}$ can be derived in a general context from $(1/\mu_0)\int \vec{B}_1\cdot\vec{B}_2\,\mathrm{d}^3\vec{r}$. That formula computes all the energy involved in creating the dipole, keeping it permanent and placing it in the external magnetic field, whereas $-\vec{m}\cdot\vec{B}^\mathrm{(ext)}$ only accounts  for the latter. For more details we  suggest consulting section 5.16 of ref. \cite{Jackson}, section 4.3 of ref. \cite{Wald} or section 12.7 of ref. \cite{Zangwill}.} 
Once again, the contribution stemming from the delta-function term in the magnetic field of the dipole was crucial for the matching of the two results.

\section{Final remarks}

Evidently, point dipoles (electric or magnetic) constitute an idealization, a fact which is partly reflected in the delta-function terms  appearing in the expressions for their fields. Nevertheless, there are several electromagnetic phenomena whose description is immensely simplified if one considers matter as being composed of (possibly time-varying) dipoles. Therefore it pays off to deal with those singular delta-function terms in the fields.

It is well know that the delta-function terms play an important role in quantum mechanics, where they are responsible for interactions between particles located at the same position (sometimes referred to as contact terms). Classically, on the other hand, we know that interactions involve sources at spatially separated space-points, which might suggest that one can always ignore the delta-function terms in classical settings.

However, if one is interested in calculating the interaction energy of a classical electromagnetic system in the most fundamental way (from a relativistic point of view \cite{Wald}), then one needs to integrate the electric and magnetic fields over space, and those delta-functions will in general give a finite contribution to the energy.

In this work, we have explicitly computed that contribution to the energy stemming from the delta-function terms in two simplified, purely classical settings. Namely an electric dipole in the center of a spherical cavity inside a conductor (an electrostatics problem) and its magnetostatic analogue, i.e. a magnetic dipole in the center of a spherical cavity inside a superconductor. In both cases, it can be clearly seen that one needs to keep the delta-function terms in order to get the correct result for the interaction energy, which can also be computed by other means.


We have focused on dipoles not only because they are simpler to deal with mathematically, but also because they are relevant for many applications. Nonetheless, the higher multipoles also contain terms involving delta functions (see, for example, the discussion in ref. \cite{multi}), and those terms should contribute to the energy as well. We leave that investigation for future work.

\section*{Acknowledgements}

I would like to thank C. Farina for helpful comments.


\begin{thebibliography}{99}


\bibitem{Griffiths} D.J. Griffiths, Introduction to Electrodynamics 4th edn (Engle-wood Cliffs, NJ: Prentice-Hall 2012).

\bibitem{Jackson} J.D. Jackson, Classical Electrodynamics 3rd edn (Hoboken,NJ: John Wiley \& Sons 1999).

\bibitem{Zangwill} A. Zangwill, Modern Electrodynamics 1st edn (New York, NY:Cambridge University Press 2012).

\bibitem{Taddei} M.M. Taddei et al., Eur. J. Phys. {\bf 31} (2010) 89.

\bibitem{Griffiths2} D.J. Griffiths, Am. J. Phys. {\bf 50}  (1982) 698.

\bibitem{Farina} Y. Muniz, A. Fonseca and C. Farina, Revista Mexicana de F\'isica E {\bf 65} (2019) 71. 

\bibitem{Wald} R. Wald, Advanced Classical Electromagnetism (Princeton, NJ: Princeton University Press 2022).

\bibitem{multi} E. Parker, Eur. J. Phys. {\bf 38}  (2017) 025205.


\end{thebibliography}
\end{document}